\documentstyle[12pt]{article}
\begin{document}
\newcommand{\vs}{\vspace{2mm}}
\newcommand{\mup}{{\mu}^{\prime}}
\newcommand{\nup}{{\nu}^{\prime}}
\newcommand{\muz}{{\mu}^{\prime \prime}}
\newcommand{\nuz}{{\nu}^{\prime \prime}}
\newcommand{\z}{\prime \prime}
\newcommand{\p}{\prime}
\begin{titlepage}
\vspace{-10mm}
\begin{flushright}
             December 2001\\ IUT-Phys/01-11\\
\end{flushright}
\vspace{12pt}
\begin{center}
\begin{large}
{\bf Dirac Quantization of Some Singular Theories}\\
\end{large}
\vspace{5mm}
{\bf A. Shirzad
\footnote{e-mail: shirzad@theory.ipm.ac.ir}}
{\bf , P. Moyassari}\\
\vspace{12pt}
{\it Department of  Physics, Isfahan University of Technology \\
Isfahan,  IRAN, \\
Institute for Studies in Theoretical Physics and Mathematics\\
P. O. Box: 5746, Tehran, 19395, IRAN.}
\vspace{0.3cm}
\end{center}
\abstract{Analyzing the constraint structure of electrodynamics, massive vector bosons, Dirac fermions and electrodynamics  coupled to fermions, we show that Dirac quantization method leads to appropriate creation-annihilation algebra among the Forier coefficients of the fields. 
 
\
\vfill}
\end{titlepage}

\section{Introduction}
The standard procedure for canonical quantization of fields is based on the rule
\begin{equation}
\{\hspace{1mm} ,\hspace{1mm}\}_{P.B}\rightarrow \frac{1}{i\hbar} [\hspace{1mm},\hspace{1mm}] \label{a1}
\end{equation}
For simple examples such as Klien-Gordon field, the above rule leads to
creation-annihilation algebra between coefficients of Forier modes
of the fields. However, the problem is not so simple in most cases
of interest such as gauge theories. In fact a large number of
models come from singular Lagrangians, where a well-defined Poisson
bracket is far from reach. In these cases, the traditional method for
quantization is as follows: One finds first the {\it physical fields} in 
some sense and considers their Forier transformations; and then  {\it assumes}
 that coefficients of
Forier modes act as creation-annihilation operators \cite{msh,ryder}. Usually, there exist 
difficulties in recognizing the physical fields such that a standard method is
not followed by authors to find them. On the other hand, no 
justifiable reason is given for assuming the creation-annihilation algebra among coefficients of Forier expansion. Accordingly we think that
this approach violates the unified quantization method given by
(\ref {a1}). However, it is well-known that a well-defined
bracket, known as Dirac bracket, can be assigned to singular
Lagrangians. In this method the quantization rule (\ref {a1}) is
generalized to
\begin{equation}
\left\{\hspace{1mm},\hspace{1mm} \right\}_{D.B} \longrightarrow \frac{1}{i\hbar} [\hspace{1mm} ,\hspace{1mm}].\label{a5}
\end{equation}
This rule reduces simply to (\ref {a1})  for regular models.

As far as we know, this unified method for quantization, i.e. Dirac
 quantization, is not exactly followed in the literature for models of physicists' interest.

Our main objective in this paper is that the appropriate
creation-annihilation algebra would emerge directly by following
the Dirac method, with no need to any ad-hoc assumption. After a
brief review of Dirac quantization in the next section, we will do
this job for electrodynamics, massive vector
bosons, Dirac fermion and electrodynamics coupled to fermions in sections 3-6.

\section {Dirac quantization}
The Lagrangian $L(q,\dot{q})$ is said to be singular if the momenta
\begin{equation}
p_{i}=\frac{\partial L}{\partial\dot{q_{i}}}\label{a6}
\end{equation}
are not independent functions of $(q,\dot{q})$. It follows that a number of functions
 $\phi_i (p,q)$ should vanish as primary constraints.
  It can be shown \cite{dirac} that time development of any function $g(q,p)$
is achieved via the equation
\begin{equation}
\dot{g}=\left\{g,H_T \right\}
\label{a7}
\end{equation}
where
\begin{equation}
H_T =H_C+\lambda^i\phi_i\label{a10}
\end{equation}
in which $H_C$ is the canonical Hamiltonian and $\lambda^i$ are Lagrange multipliers.
 Primary constraints should vanish at all times. So the consistency conditions
\begin{equation}
\dot{\phi}_i=\left\{\phi_i,H_C\right\}+\lambda^{j}\left\{\phi_i,\phi_{j}\right\}\approx 0 \label{a8}
\end{equation}
should be imposed on them, where $\approx$ means weak equality,
i.e. equality on the constraint surface. The consistency conditions
(\ref{a8}) may lead to determining a number of Lagrange multipliers
and/or emerging some new constraints, known as secondary
constraints. The procedure should be repeated until the last step
where the maximum number of Lagrange multipliers are determined
and the consistency of all remaining constraints are guaranteed on
the constraint surface.

The whole system of constraints divide to
first class constraints $\phi_a$ and second class constraints
$\chi_n$ such that
\begin{equation}
\begin{array}{l}
\left\{\phi_a,\phi_{a'}\right\}\approx 0 \vs \\ \left\{ \phi_a,\chi_n\right\} \approx 0 \\ \vs
\left\{\chi_n,\chi_m\right\}\approx c_{nm}    \hspace{1.5cm} det(c_{nm}\neq0).\\
\end{array}
\label{a9}
\end{equation}
The second class constrains, which are even in number, reduce the
dimensionality of phase space and lead to determining  a number of
the Lagrange multipliers \cite{dirac}. While, the first class constraints are
generators of gauge transformations \cite{grapo}. In order to fix the gauges one
should impose a number of gauge fixing conditions
$\Omega_{a'}(q,p)=0$ on the system. For gauges to be fixed
completely, the determinant of $\left\{\Omega_{a'},\phi_{a}\right\}$
should not vanish \cite{shl}. In other words, after fixing the gauges the set of
constraints and gauge fixing conditions, i.e.
\begin{equation}
\Phi_A\equiv(\chi_n;\phi_a,\Omega_{a'}),\label{a11}
\end{equation}
should act as a system of second class constraints. The reduced phase
space $\Phi_A\approx0$ possesses a well-defined bracket, known as
Dirac bracket, which is defined for any two arbitrary functions
$f(q,p)$ and $g(q,p)$ as
\begin{equation}
\left\{f,g\right\}_{D.B}=\left\{f,g\right\}_{P.B}-\left\{f,\Phi_A\right\}_{P.B}C^{AB}\left\{\Phi_B,g\right\}_{P.B}\label{a12}
\end{equation}
where $C^{AB}$ is the inverse of
$C_{AB}=\left\{\Phi_A,\Phi_B\right\}_{P.B}$. In this way all redundant
coordinates are omitted  and just the physical degrees of
freedom are remained in the reduced phase space. Then one can
quantize the system in a consistent way by  transforming the Dirac brackets into 
commutators, as mentioned before in (\ref{a5}). Let us now
 follow the procedure for some important physical models.
\section {Electrodynamics}
The Lagrangian density for electrodynamics reads
\begin{equation}
{\cal L}={\textstyle -\frac{1}{4}}F^{\mu\nu}F_{\mu \nu} \label{a12}
\end{equation}
where
\begin{equation}
F_{\mu\nu}=\partial_{\mu} A_{\nu}-\partial_{\nu} A_{\mu}. \label{a13}
\end{equation}
From definition of momentum fields, i.e.
\begin{equation}
\Pi ^{\mu}(x)=\frac{\delta L}{\delta(\partial_0 A_{\mu}(x))}=-F^{0 \mu}(x), \label{a14}
\end{equation}
one finds that $\Pi^0 (x)$ vanishes. So the primary constraint of the system is
\begin{equation}
\phi^{1}(x)=\Pi^0(x). \label{a15}
\end{equation}
To quantize the theory most authors replace the main Lagrangian with Fermi Lagrangian:  
\begin{equation}
{\cal L}_{\mbox f}={\textstyle -\frac{1}{2}}(\partial_{\mu}A_{\nu})(\partial^{\mu} A^{\nu})    \label{a16}
\end{equation}
which, together with the Loranz condition $\partial_{\mu} A^{\mu}=0$,
gives the same equations of motion \cite{msh,itz}. It is well-known that
the Fermi Lagrangian leads to negative norm states after
quantization, first problem; and the Loranz gauge condition can
not be imposed in the operator form, next problem. To this end,
one can show that imposing the condition
\begin{equation}
\langle{\rm phy}|\partial_{\mu} A^{\mu}|{\rm phy}\rangle =0\label{a17}
\end{equation}
on physical states solves both problems. However, it seems that the
above method is adjusted just for the problem of quantization of
electrodynamics. While it is preferable to follow a general
procedure to quantize all models.

Now we show that following the Dirac method, one can quantize the
theory in a straightforward method. Returning to the main Lagrangian(\ref{a12}),
the canonical Hamiltonian reads
\begin{equation}
H_C=\int \left({\textstyle \frac{1}{2}}\Pi_i\Pi^i+A_0\partial_i \Pi^i +{\textstyle\frac{1}{4}}F_{ij}F^{ij}\right)d^3x\label{a18}
\end{equation}
Consistency of $\phi^1$ from (\ref{a8}) gives the second level constraint
\begin{equation}
\phi^2\equiv\left\{\phi^1 ,H_C\right\}=\partial_i\Pi^i \label{a19}
\end{equation}
It is easy to see that $\left\{\phi^2 ,\phi^1\right\}\approx 0 $ and $\left\{\phi^2 ,H_C\right\}\approx 0 $.
Therefore, there is only one constraint chain as
\begin{equation}
\begin{array}{l}
\phi^1=\Pi^0 \vs \\ \phi^2=\partial_i \Pi^i
\end{array}
\label{a20}
\end{equation}
As discussed in \cite{shl}, to fix the gauge it is enough to
impose the gauge fixing condition
\begin{equation}
\Omega^2=\partial_i A^i . \label{a21}
\end{equation}
Since $\left\{\Omega^2,\phi^2\right\} \neq0 $, $\Omega^2 $  fixes the gauge transformation generated by $\phi^2$.
Consistency of gauge fixing condition $\Omega^2 $ gives the other
gauge fixing condition conjugate to $ \phi^1 $:
\begin{equation}
\Omega^1\equiv \left\{\Omega^2 ,H_T\right\}\approx \partial^i \partial_i A_0 \label{a23}
\end{equation}
Finally, one should calculate the Dirac brackets due to system of constraints
\begin{equation}
(\phi^1,\phi^2;\Omega^2,\Omega^1) . \label{a24}
\end{equation}
After a little algebra, the Dirac brackets emerges as
\begin{equation}
\begin{array}{l}
{\displaystyle \left[A^{\mu}({\bf x},t),\Pi_{\nu}({\bf x'},t)\right]_{D.B}=(\delta^{\mu}_{\ \ \nu}   -g_{\nu0}\delta^{\mu}_{\ \ 0})\delta({\bf x}-{\bf x'})-\partial^{\mu}\partial_{\nu} \frac{1}{4\pi|{\bf x}-{\bf x}'|} } \vs \\
\vs \left[A^{\mu}({\bf x},t),A_{\nu}({\bf x}',t)\right]_{D.B}=0  \vs \\  \left[\Pi^{\mu}({\bf x},t),\Pi_{\nu}({\bf x}',t)\right]_{D.B}=0 
\end{array}
\label{a25}
\end{equation}
Assuming the system to be contained in a box of volume $V$, the
most general solution to equations of motion reads
\begin{equation}
A^{\mu}(x)=\sum_k \sum^3_{r=0}  \frac{1}{(2Vw_k)^{1/2}}(\epsilon^{\mu}_{r}({\bf k})a_r({\bf k})e^{-ikx}+\epsilon^{\mu}_{r}({\bf k}) a^\dagger_r({\bf k})e^{ikx})\label{a26}
\end{equation}
Where $w_k=|{\bf k}|    $ and the polarization vectors $\epsilon^{\mu}_r({\bf k})   $ can be chosen to be
\begin{equation}
\begin{array}{ll}
{\rm scalar:}&\epsilon^{\mu}_{0}({\bf k})=(1,0,0,0) \vs \\ {\rm
transverse:}&\epsilon^{\mu}_{1,2}({\bf k})=(0,{\bf e}({\bf k}))
\hspace{0.5cm}\ {\bf k}.{\bf e}({\bf k})=0  \vs \\{\rm
longitudinal:}&\epsilon^{\mu}_{3}({\bf k})=(0,\frac{{\bf k}}{|{\bf k}|})
\end{array}
\label{a27}
\end{equation}
Using (\ref{a14}), the momentum fields can also be written as
\begin{equation}
\Pi^{\mu}(x)=\sum_k \sum^3_{r=0}  \frac{i}{(2Vw_k)^{1/2}}\left(k^{\mu}\epsilon^{0}_{r}({\bf k})-w_k\epsilon^{\mu}_{r}({\bf k})\right) \left(-a_r({\bf k})e^{-ikx}+a^\dagger_r({\bf k})e^{ikx}\right) \label{a28}
\end{equation}
Using (\ref{a26}) and   (\ref{a28}), one can find $a_r({\bf k})$ in term of fields (at $t=0$) as
\begin{equation}
\xi_ra_r({\bf k})=\frac{1}{(2w_k)^{1/2}} \int d^3x \epsilon_{\mu r} \left(-i\Pi^{\mu}-w_kA^{\mu}\right) e^{i{\bf k}.{\bf x}}+\frac{k^j \epsilon_{jr}}{w_k} a_0({\bf k}) \label{a29}
\end{equation}
where $\xi_0=-1    $ and $\xi_{1,2,3}=1     $. Finally, from
(\ref{a25}), (\ref{a29}) and its complex conjugate for $a^\dagger_r({\bf k}) $
one can derive the Dirac brackets among $a_r({\bf k})$ and $a^\dagger_r({\bf k})
$ as:
\begin{equation}
\begin{array}{l}
\left[a_0({\bf k}),a^\dagger_0({\bf k}')\right]_{D.B}=0 \vs \\
 \left[a_3({\bf k}),a^\dagger_3({\bf k}')\right]_{D.B}=0 \vs \\ \left[a_r({\bf k}),a^\dagger_s({\bf k}')\right]_{D.B}=-i \delta_{rs} \delta_{k,k'}  \hspace{0.5cm} 
 r,s=1,2 \\
\end{array}
\label{a30}
\end{equation}
Now everything is ready for quantization. It is just needed to
use quantization rule (\ref{a5}) to find commutators. As is apparent the
 scalar and longitudinal polarizations
disappear in a consistent way, and 
$a_{1,2}({\bf k})$ and $a^\dagger_{1,2}({\bf k})$ obey creation-annihilation algebra. 
In other words, just transverse photons remain as physical degrees of
 freedom in the quantized theory.  
\section{Massive Vector Bosons}
The model is given by the Lagrangian density
\begin{equation}
{\cal L}={\textstyle-\frac{1}{2}}F^{\dagger}_{\mu\nu}F^{\mu\nu}+m^2W^{\dagger\mu}W_{\mu} \label{a31}
\end{equation}
where the complex fields $W^\mu$ and $W^{\mu \dagger}$ describe vector bosons of mass $m$ and
\begin{equation}
F_{\mu\nu}=\partial_\mu W_\nu -\partial_\nu W_\mu . \label{a32}
\end{equation}
The equations of motion reads
\begin{equation}
\partial_\mu F^{\mu\nu}+m^2W^{\nu}=0 . \label{a33}
\end{equation}
(similar relations as (\ref{a32}) and (\ref{a33}) can be understood for $W^{\dagger\mu}$ and 
$F^{\dagger \mu \nu}$). Taking the divergence of both sides of (\ref{a33}) gives
\begin{equation}
\partial_\nu W^\nu =0 \label{a34}
\end{equation}
In the context of the theory of constrained systems  $\partial_\nu
W^\nu    $ (as well as $\partial_\nu W^{\dagger\nu} $) is recognized as a
Lagrangian constraint \cite{shirzad}. It shows that the fields $W^\mu$ are not independent.
Traditionally, people consider the scalar polarization
as the redundant field and quantize the remaining
ones (i.e. transverse and Longitudinal polarizations) as
physical degrees of freedom. But usually there is no justification that "why
scalar polarization should be omitted?". Now, using the theory of
constrained systems, we show that this is really the case.
From (\ref{a31}) 
the canonical momentum fields are $\Pi^ \mu =F^{\dagger\mu0} $ and $\Pi^ {\dagger\mu }=F^{\mu0}  $. This  leads to primary constraints
\begin{equation}
\phi^{1}_{1}=\Pi^0 \hspace{0.5cm} \phi^{1}_{2}=\Pi^{\dagger0} \label{a35}
\end{equation}
The canonical Hamiltonian reads
\begin{equation}
H_C=\int d^3x \left(\Pi^\dagger_i \Pi^i +\partial_i \Pi^i W_0 + \partial_i \Pi^{\dagger i} W^\dagger_0 +
{\textstyle \frac{1}{2}}F^{ij} F^\dagger_{ij}-m^2 W^\dagger_\mu W^\mu \right)
\label{a36}
\end{equation}
Consistency conditions (\ref{a8}) on primary constraints gives the
second level constraints as
\begin{equation}
\phi^{2}_{1}=\partial_i\Pi^i+m^2W^{\dagger}_0 \hspace{1.2cm} \phi^{2}_{2}=\partial_i\Pi^{\dagger i}+m^2W_0 \\
\label{a37}
\end{equation}
It is easy to see that we have four second class constrains. In
fact since $\left\{\phi^{2}_{1},\phi^{1}_{2}\right\} \neq0$
and $\left\{\phi^{2}_{2},\phi^{1}_{1}\right\} \neq0$  we have
two {\it cross conjugate} second class chains, in the terminology
of \cite{lsh}. The Dirac brackets can be derived straightforwardly
as
\begin{equation}
\begin{array}{l}
{\displaystyle \left[W_i({\bf x},t),\Pi^j({\bf x}',t)\right]_{D.B}=\left[W^\dagger_i({\bf x},t),\Pi^{\dagger j}(x',t)\right]_{D.B}=\delta_i^j \delta({\bf x}-{\bf x'}) }
 \vs \\ {\displaystyle \left[W_0({\bf x},t),\Pi^0({\bf x}',t)\right]_{D.B}=\left[W^\dagger_0({\bf x},t),\Pi^{\dagger 0} ({\bf x}',t)\right]_{D.B}=0 }
\end{array}
\label{a38}
\end{equation}
As in electrodynamics, the solution of equations of motion reads
\begin{equation}
W^{\mu}(x)=\sum_k \sum^3_{r=0}  \frac{1}{(2Vw_k)^{1/2}}\left[\epsilon^{\mu}_r({\bf k})a_r({\bf k})e^{-ikx}+\epsilon^{\mu}_r({\bf k}) b^\dagger_r({\bf k})e^{ikx}\right]\label{a39}
\end{equation}
where $w_k=\sqrt{k^2+m^2}$. Using (\ref{a39}), its complex conjugate for $W^\dagger_\mu$,
 and expression for $\Pi_ \mu$ and $\Pi^\dagger_\mu$, one finds
\begin{equation}
\begin{array}{l}
{\displaystyle \xi_r a_r=\frac{1}{(2w_k)^{1/2}} \int d^3x\epsilon_{\mu r}({\bf k})(-i\Pi^{\dagger\mu}({\bf x})-w_kW^\mu ({\bf x}))e^{i{\bf k}.{\bf x}}+\frac{k^j\epsilon_{jr}({\bf k})}{w_k}a_0({\bf k}) }
 \vs \\ {\displaystyle \xi_r b_r=\frac{1}{(2w_k)^{1/2}} \int d^3x\epsilon_{\mu r}({\bf k})(-i\Pi^{\mu}({\bf x})-w_kW^{\dagger\mu}({\bf x}))e^{i{\bf k}.{\bf x}} +\frac{k^j\epsilon_{jr}({\bf k})}{w_k}b_0({\bf k})}
\end{array}
\label{a40}
\end{equation}
where the fields are considered at $t=0$. The expression for $a^\dagger_r({\bf k})  $
 and $b^\dagger_r({\bf k})  $ can be written by complex conjugating the above formulas. 
Then using the Dirac brackets (\ref{a38}) the following fundamental Dirac berackets emerge
\begin{equation}
\begin{array}{l}
{\displaystyle\left[a_r ({\bf k}),a^\dagger_s ({\bf k}') \right]_{D.B}=-i\delta_{rs}\delta_{kk'} \hspace{1cm} r,s=1,2,3 }\vs 
\\ \left[b_r ({\bf k}),b^\dagger_s ({\bf k}') \right]_{D.B}=-i\delta_{rs}\delta_{kk'} \hspace{1cm} r,s=1,2,3  \vs 
\\ \left[a_0 ({\bf k}),a^\dagger_0 ({\bf k}') \right]_{D.B}=0  \vs 
\\ \left[b_0 ({\bf k}),b^\dagger_0 ({\bf k}') \right]_{D.B}=0 
\end{array}
\label{b41}
\end{equation}
As is apparent, after quantization  according to (\ref{a5}), the
scalar polarization disappears. Moreover $a_r({\bf k})$ and
$a^\dagger_r({\bf k})$ act as annihilation  and creation operators for
$W^+$ particle with transverse and longitudinal
polarizations, while $b_r({\bf k})$ and $b^\dagger_r({\bf k})$ act  in the same
way for $W^-$ particle.
\section{Dirac Fermions}
Fermion fields satisfy different algebra, compared to bosons, even at the classical level.
In fact the Poisson bracket of two functions on a super phase space have generalized algebraic
properties \cite{heno}, among them we remind the following:
\begin{equation}
\begin{array}{l}
\left\{ F,G \right\}=(-1)^{\epsilon_F \epsilon_G+1}\left\{ G,F \right\} \vs \\ 
\left\{F,GH \right\}=\left\{F,G\right\}H+(-1)^{\epsilon_F \epsilon_G} G  \left\{F,H\right\}
\end{array}
\label{a41}
\end{equation}
where $\epsilon_F$ and $\epsilon_G$ are Grassmanian parity of $F$ and $G$ respectively.
This shows specially that the Poisson bracket of two fermionic variables  is even under changing the
order. As we will show, this leads to defining anticommutators, instead of commutators,
 for the operators at the quantum level.
The Dirac fields are introduced within the Lagrangian density
\begin{equation}
{\cal L}=\bar{\Psi}(x)(i\gamma^\mu \partial_\mu -m)\Psi(x) \label{a42}
\end{equation}
where $\Psi(x)$ and $\bar{\Psi}(x)$ are fermionic fields. The canonical momentum fields are derived as
\begin{equation}
\begin{array}{l}
{\displaystyle \Pi(x)=\frac{\partial^L L}{\partial\dot{\Psi}}=-i\bar{\Psi}\gamma^0 }  \vspace{5mm} \\ 
{\displaystyle \bar{\Pi}(x)=\frac{\partial^L L}{\partial\dot{\bar{\Psi}}}=0 }
\end{array}
\label{a43}
\end{equation}
where   $\partial^L$ means left differentiation. From (\ref{a43}) the following primary constrains
emerge
\begin{equation}
\phi_1=\Pi+i\bar{\Psi}\gamma^0 \hspace{1cm} \phi_2= \bar{\Pi}
\label{a44}
\end{equation}
As is obvious, $\{ \phi_1,\phi_2 \} \neq 0 $, showing that $\phi_1$ and $\phi_2$ are second class. Therefore, the consistency condition on them just determines the
 Lagrang multipliers. Using the fundamental Poisson brackets
 of fermion fields, i.e.
\begin{equation}
\left\{ \Psi({\bf x},t),\Pi({\bf x}',t) \right\}=\left\{ \bar{\Psi}({\bf x},t),\bar{\Pi}({\bf x}',t) \right\}=-\delta({\bf x}-{\bf x'}), \label{b1}
\end{equation} 
the Dirac berackets, with the same
algebraic properties as Poisson bracket, are simply derived as
\begin{equation}
\begin{array}{l}
{\displaystyle \left[\Psi ({\bf x},t),\bar{\Psi} ({\bf x}',t)\right]_{D.B} =-i\gamma^0 \delta^3
(x-x') }\vs \\\left[\Psi ({\bf x},t),\Pi ({\bf x}',t)\right]_{D.B} =-\delta^3 (x-x') \vs \\
\left[\bar{\Psi}({\bf x},t),\bar{\Pi}({\bf x}',t)\right]_{D.B}=0 \vs \\
\left[\Pi({\bf x},t),\bar{\Pi}({\bf x}',t)\right]_{D.B}=0 
\end{array}
\label{a45}
\end{equation}
The famous Forier expansion of Dirac fields, read
\begin{equation}
\begin{array}{l}
{\displaystyle \Psi(x)=\sum_k \sum^2_{r=1}(\frac{m}{VE_p})^{1/2}\left[c_r({\bf p})u_r({\bf p})e^{-ipx}+d^\dagger_r({\bf p})v_r({\bf p})e^{ipx}\right] } \vs \\
{\displaystyle \bar{\Psi}(x)=\sum_k \sum^2_{r=1}(\frac{m}{VE_p})^{1/2}\left[c^\dagger_r({\bf p})\bar{u}_r({\bf p})e^{ipx}+d_r({\bf p})\bar{v}_r({\bf p})e^{-ipx}\right] }
\end{array}
\label{a46}
\end{equation}
where $u_r({\bf p})$ and $v_r({\bf p}) $ are Dirac spinors. At this point text
books on quantum field theory usually {\it assume} the following  anti-commutator algebra between Forier coefficients
as the beginning point for quantization of Dirac fermions
\begin{equation}
\begin{array}{l}
\left[c_r({\bf p}) ,c^\dagger _s({\bf p}')\right]_+=\delta_{rs} \delta_{pp'} \vs \\
 \left[d_r({\bf p}) ,d^\dagger _s({\bf p}')\right]_+=\delta_{rs}\delta_{pp'}
\end{array}
\label{a47}
\end{equation}
However, using
\begin{equation}
\begin{array}{l}
{\displaystyle 2\sqrt{m/E_p} \: c_r({\bf p})=\int d^3x \bar{u}_r({\bf p}) \Psi({\bf x},0)e^{i{\bf p}.{\bf x}} }  \vs
 \\ {\displaystyle -2\sqrt{m/E_p} \: d_r({\bf p})=\int d^3x \bar{\Psi}({\bf x},0)v_r({\bf p}) e^{i{\bf p}.{\bf x}} }
\end{array}
\label{a48}
\end{equation}
and (\ref {a45}) it is easy to see that
\begin{equation}
\begin{array}{l}
\left[c_r({\bf p}),c^\dagger_s({\bf p}')\right]_{D.B}=-i\delta_{rs}\delta_{pp'} \vs \\
\left[d_r({\bf p}),d^\dagger_s({\bf p}')\right]_{D.B}=-i\delta_{rs}\delta_{pp'}
\end{array}
\label{a49}
\end{equation}
Note that the above (classical) berackets are even under exchange of variables. It is reasonable to keep
the algebraic properties of classical berackets at quantum level. So in the quantization
rule (\ref{a5}) we should replace anti-commutators on the right hand side, in case of fermions.
 The final result is the creation -annihilation
algebra(\ref{a47}). We see once again that Dirac method for quantization
 gives the correct algebra of
quantum operators with no need to unjustified assumptions.

\section{Electrodynamics coupled to Fermions}
The Lagrangian density for electromagnetic field coupled to Dirac fermions is known as
\begin{equation}
{\cal L}=-{\textstyle\frac{1}{4}}F^{\mu\nu}F_{\mu\nu}+\bar{\Psi}(i\gamma^\mu\partial_\mu-m)\Psi-e\bar{\Psi}\gamma^\mu A_\mu\Psi \label{a50}
\end{equation}
The system of constraints can be obtained similar to the
 case of free electrodynamics and Dirac fermions as follows:
\begin{equation}
\begin{array}{lll}
{\displaystyle \phi^1_1=\Pi^0 \hspace {5cm} \phi_2=\bar{\Pi} \hspace{2cm} \phi_3=\Pi+i\bar{\Psi}\gamma^0 \vspace{0.3cm} }\vs \\
{\displaystyle \phi^2_1=-\partial_i\Pi^i+ie(\bar{\Psi}\bar{\Pi}+\Pi\Psi) }
\end{array}
\label{a51}
\end{equation}
where $\phi^2_1=\left\{\phi^1_1,H_T\right\}$. As before, $\phi_2$ and $\phi_3$ are second class; however, $\phi_1^1$ and $ \phi^2_1$ are
elements of a first class chain. Parenthetically, the infinitesimal gauge transformation
\begin{equation}
\begin{array}{l}
A_\mu \longrightarrow A_\mu+\partial_\mu\zeta \vs \\ 
\Psi\longrightarrow(1-ie\zeta)\Psi \vs \\ 
\bar{\Psi}\longrightarrow(1+ie\zeta)\bar{\Psi},
\end{array}
\label{b2}
\end{equation}
can be obtained as the action of the gauge generator \cite{grapo,shisha}
\begin{equation}
G=\int d^3x(\dot{\zeta}\phi^1_1-\zeta\phi^2_1). \label{a52}
\end{equation}
In order to fix the gauge, following the method of \cite{shl}, we can introduce the gauge fixing
condition
\begin{equation}
\Omega^2_1=\partial_iA^i \label{a53}
\end{equation}
conjugate to the last element $\phi^2_1$.
Then consistency of $\Omega^2_1$ gives the next gauge fixing condition as
\begin{equation}
\Omega^1_1=\partial_i\partial^i A_0 \label{a54}
\end{equation}
The whole system of constraints $\left(\phi^1_1,\phi^2_1,\Omega^2_1,\Omega^1_1;\phi_2,\phi_3\right)$ imply the following Dirac berackets
\begin{equation}
\begin{array}{l}
{\displaystyle\left[A^{\mu}({\bf x},t),\Pi_{\nu}({\bf x}',t)\right]_{D.B}=(\delta^{\mu}_{\ \ \nu}   -g_{\nu0}\delta^{\mu}_{\ \ 0})\delta({\bf x}-{\bf x'})-\partial^{\mu}\partial_{\nu} \frac{1}{4 \pi |{\bf x}-{\bf x}'|} } \vs 
\\ \left[\Psi({\bf x},t),\bar{\Psi}({\bf x}',t)\right]_{D.B}=-i\gamma^0\delta({\bf x}-{\bf x'}) \vs 
\\ \left[\Psi({\bf x},t),\Pi({\bf x}',t)\right]_{D.B}=-\delta({\bf x}-{\bf x'}) .
\end{array}
\label{a55}
\end{equation}
All other Dirac brackets vanish. These are the same brackets as obtained for free electrodynamics 
and free Dirac fermions in (\ref{a25}) and
(\ref{a45}). This shows that the algebra of quantum fields are the same as before. But some care is
necessary to obtain the appropriate algebra among the coefficients of Forier expansion. That is because in the presence
of coupling term $-e\bar{\Psi}\gamma^\mu A_\mu\Psi$ the equations of motion are different from that
for free fields and the Forier expansions (\ref{a26}) and (\ref {a46}) are no longer valid.
However, in quantum theory the coupling term $-e\bar{\Psi}\gamma^\mu A_\mu\Psi$ is
considered as a perturbation. So the quantum fields obey the same equations of motion as obtained
from the free part of the Lagrangian (i.e. the first two terms in (\ref{a50})). Consequently the fields
$A^\mu$, $\Psi$ and $\bar{\Psi}$ have the same Forier expansions as in the free cases. To this end, beginning
with quantum versions of (\ref{a55}) one is allowed to use free field expansion (\ref{a26}) and (\ref{a46}) to
obtain the same creation-annihilation algebra as in (\ref{a30}) and (\ref{a49}).

\end{document}